\documentclass[proceedings,opts]{JHEP}

\newcommand{\be}{\begin{eqnarray}}
\newcommand{\ee}{\end{eqnarray}}
\newcommand{\non}{\nonumber}
\newcommand{\n}{\ensuremath{\mathcal{N}}}
\newcommand{\tr}{\mathop{\rm tr}\nolimits}

\conference{TMR Non perurbative quantum effects 2000}

\title{``New'' boundary conditions in integrable lattice models}

\author{Anastasia Doikou\\ Department of Mathematical Sciences, University of Durham, DH1 3LE Durham, England\\ Anastasia.Doikou@durham.ac.uk}

\abstract{We consider the case of an integrable quantum spin chain with ``soliton non-preserving'' boundary conditions. This is the first time that such boundary conditions have been considered in the spin chain framework. We construct the transfer matrix of the model, we study its symmetry and we find explicit expressions for its eigenvalues. Moreover, we derive a new set of Bethe ansatz equations by means of the analytical Bethe ansatz method.} 

\begin{document}

\section{Introduction}

 So far, quantum spin chains with ``soliton preserving'' boundary conditions

have been studied \cite{DVGR3}-\cite{dn/duality}. However, there exists another type of boundary conditions, namely the ``soliton non-preserving''. These conditions are known in affine Toda field theories \cite{delius1}-\cite{gand}, although there is already a hint of such boundary conditions in the prototype paper of Sklyanin \cite{sklyanin}, which is further clarified by Delius in \cite{delius1}. It is important to mention that in affine Toda field theories only the ``soliton non-preserving'' boundary conditions have been studied \cite{gand}, \cite{corr}. It is still an open question what the ``soliton preserving'' boundary conditions are in these theories.

In this work we construct the open spin chain with the ``new'' boundary conditions, we show that the model is integrable, we study its symmetry, and evidently, we solve it by means of the analytical Bethe ansatz method \cite{reshe}-\cite{amn/spectrum}. This is the first time that such boundary conditions have been considered in the spin chain framework.

To describe the model it is necessary to introduce the basic constructing elements, namely, the $R$ and $K$ matrices.

 The $R$ matrix, which is a solution of the Yang-Baxter equation
\be
R_{12}(\lambda_{1} - \lambda_{2})\ R_{1 3}(\lambda_{1})\ R_{23}(\lambda_{2}) \non\\ = R_{23}(\lambda_{2})\ R_{1 3}(\lambda_{1})\ R_{1 2}(\lambda_{1} - \lambda_{2})\
\label{YBE}
\ee
(see, e.g., \cite{QISM}). 

Here, we focus on the special case of the $SU(3)$ invariant $R$ matrix
\cite{yang}
\be
R_{12}(\lambda)_{j j \,, j j} &=& {(\lambda + i)} \,, \non \\
R_{12}(\lambda)_{j k \,, j k} &=& { \lambda} \,, \qquad j \ne k
\,, \non \\
R_{12}(\lambda)_{j k \,, k j} &=& {i}\,, \qquad j \ne k \,, \non \\
& & 1 \le j \,, k \le 3 \,.
\label{Rmatrix}
\ee
We also need to introduce the $R$ matrix that involves different representations of $SU(3)$ \cite{VEWO}, \cite{abad}, in particular, $3$ and $\bar 3$ (see also \cite{doikou}). This matrix is given by crossing \cite{ogiev}-\cite{delius}
\be
R_{\bar 12}(\lambda) = V_{1}\ R_{12}(-\lambda - \rho)^{t_{2}}\ V_{1} \non\\
= V_{2}^{t_{2}}\ R_{12}(-\lambda - \rho)^{t_{1}}\ V_{2}^{t_{2}} \,,
\label{prop4}
\ee
where $V^{2}=1$. 
$R_{\bar1 2}(\lambda)$ is also a solution of  the Yang-Baxter equation
\be
R_{\bar 1 2}(\lambda_{1} - \lambda_{2})\ R_{\bar1 3}(\lambda_{1})\ R_{23}(\lambda_{2}) \non\\ 
= R_{23}(\lambda_{2})\ R_{\bar1 3}(\lambda_{1})\ R_{\bar1 2}(\lambda_{1} - \lambda_{2})\,.
\label{YBE2}
\ee
The matrices $K^{-}$, and $K^{+}$ which are solutions
of the boundary Yang-Baxter equation \cite{cherednik}, \cite{gand}
\be
R_{12}(\lambda_{-})\ K_{1}^{-}(\lambda_{1})\
R_{2 \bar1}(\lambda_{+})\ K_{2}^{-}(\lambda_{2}) \non \\
= K_{2}^{-}(\lambda_{2})\ R_{\bar12}(\lambda_{+})\
K_{1}^{-}(\lambda_{1})\ R_{21}(\lambda_{-}) \,,
\label{boundaryYB1}
\ee
where $\lambda_{+} =\lambda_{1}+\lambda_{2}$, $\lambda_{-} =\lambda_{1}-\lambda_{2}$, and
\be
K^{+}(\lambda) = K^{-}(-\lambda-\rho)\,,
\label{boundaryYB2}
\ee
where $\rho = {3i \over 2}$. We can consider that the $K_{i}$ matrix describes the reflection of a soliton with the boundary which comes back as an anti-soliton. 

 It is a natural choice to consider the following alternating spin chain \cite{VEWO}, \cite{abad}, which leads to a local Hamiltonian. The corresponding transfer matrix $t(\lambda)$ for the open chain of $N$ sites with ``soliton non-preserving'' boundary conditions is (see also e.g., \cite{sklyanin}, \cite{mn/nonsymmetric})
\be
t(\lambda) = \tr_{0}  K_{0}^{+}(\lambda)\
T_{0}(\lambda)\  K^{-}_{0}(\lambda)\ \hat T_{\bar 0}(\lambda)\,,
\label{transfer1}
\ee
where $\tr_{0}$ denotes trace over the ``auxiliary space'' 0,
$T_{0}(\lambda)$ is the monodromy matrix. We define for $N$ even
\be
T_{0}(\lambda) &=& R_{0N}(\lambda) R_{0 \bar N-1}(\lambda)\cdots  R_{0 \bar 1 }(\lambda) \,, \non\\ \hat T_{\bar 0}(\lambda) &=& R_{\bar 1 \bar 0}(\lambda) R_{2 \bar 0}(\lambda)\cdots  R_{N \bar 0}(\lambda) \,,
\label{hatmonodromy}
\ee
 (we usually suppress the ``quantum-space'' subscripts
$1 \,, \ldots \,, N$).
The transfer matrix satisfies the commutativity property
\be
\left[ t(\lambda)\,, t(\lambda') \right] = 0 \,.
\label{commutativity}
\ee
We can change the auxiliary space to its conjugate and then we obtain the $\bar t(\lambda)$ matrix which satisfies, for $K^{-} = K^{+} = 1$,
\be
\bar t(\lambda) = t(\lambda)^{t}
\ee
and it also has the commutativity property,
\be
\left[ \bar t(\lambda)\,, \bar t(\lambda') \right] = 0 \,.
\label{commutativityp}
\ee
The corresponding open spin chain Hamiltonian $\cal H$ is
\be
{\cal H} \propto {d\over d \lambda}t(\lambda) \bar t(\lambda) \Big\vert_{\lambda=0} \,,
\ee
and one can show that this is indeed a local Hamiltonian with terms that describe interaction up to four neighbours.

\section{Bethe ansatz equations}

We can use the results of the previous sections in order to deduce the Bethe ansatz equations for the spin chain. First, we have to derive a reference state, namely the pseudo-vacuum.
We consider the state with all spins up i.e.,

\be
|\Lambda^{(0)} \rangle =  \bigotimes_{k=1}^{N} |+ \rangle_{(k)}\,,
\label{state}
\ee
 this is annihilated by ${\cal J}^{+}$ where (we suppress the $(k)$ index)
\be
|+ \rangle = \left (\begin{array}{c}
                     1 \\
                     0  \\ 
                     0  \\
                       \end{array} \right)\,.
\label{col}
\ee
This is an eigenstate of the transfer matrix. The action of the $R_{0k}$, $R_{\bar 0k}$ matrices on the $|+ \rangle$ ($\langle +|$) state gives upper (lower) triangular matrices.
So, the action of the monodromy matrix on the pseudo-vacuum gives also triangular matrices (see also \cite{doikou}).
We find that the transfer matrix eigenvalue for the pseudo-vacuum state, after some tedious calculations, is
\be
\Lambda^{(0)}(\lambda)&=& (a(\lambda) \bar b(\lambda)) ^{N} {2 \lambda + {i \over 2} \over 2\lambda+{3i \over 2}} + (b(\lambda) \bar b(\lambda))^{N} \non\\&+& (\bar a(\lambda)b(\lambda))^{N}{2\lambda+ {5i \over 2} \over 2\lambda+{3i \over 2}}\,.
\ee
 One can show that the model has $SO(3)$ symmetry (see \cite{last}), therefore there exist simultaneous eigenstates of $M={1\over 2}(N-S)$ and the transfer matrix, namely,
\be
M|\Lambda^{(m)} \rangle \ &=& m|\Lambda^{(m)} \rangle \,, \non\\ t(\lambda)|\Lambda^{(m)} \rangle\ &=& \Lambda^{(m)}(\lambda) |\Lambda^{(m)} \rangle\,.
\label{eigen}
\ee
We assume that a general eigenvalue has the form of a ``dressed'' pseudo-vacuum  eigenvalue i.e.,
\be
\Lambda^{(m)}(\lambda) &=&(a(\lambda) \bar b(\lambda)) ^{N} {2 \lambda + {i \over 2} \over 2\lambda+{3i \over 2}}A_{1}(\lambda) \non\\&+& (b(\lambda) \bar b(\lambda))^{N} A_{2}(\lambda)\non\\ &+& (\bar a(\lambda)b(\lambda))^{N}{ 2\lambda+ {5i \over 2} \over 2\lambda+{3i \over 2}}A_{3}(\lambda) \,.
\ee
Our task is to find explicit expressions for the $A_{i}(\lambda)$. We consider all the conditions we derived previously.
The asymptotic behaviour of the transfer matrix
\be
t(\lambda) = \lambda^{2N}(3 + {9NI \over 2 \lambda})I
\ee
gives the following condition for $\lambda \rightarrow \infty$
\be
\sum_{i=1}^{3}A_{i}(\lambda) \rightarrow 3\,.
\ee
>From  the fusion equation (see e.g., \cite{mn/fusion})
\be
\hat t(\lambda) &=& \zeta'(2\lambda+2\rho)\ \bar t(\lambda)\ t(\lambda + \rho) \non\\ &-& \zeta(\lambda+\rho)^{N/2}  \zeta'(\lambda+\rho)^{N/2} \non\\ &\times& g(2\lambda +\rho)g(-2 \lambda - 3\rho)\,, 
\label{fusiont}
\ee
where we define,
\be
g(\lambda) &=& \lambda + i\,, \qquad \zeta(\lambda) = (\lambda+i)(-\lambda+i)\,, \non\\ \zeta'(\lambda)&=&(\lambda+\rho)(-\lambda+\rho)\,,
\ee 
we obtain conditions involving $A_{1}(\lambda)$, $A_{3}(\lambda)$,
\be
A_{1}(\lambda + \rho) A_{3}(\lambda) = 1\,. 
\ee
The crossing symmetry of the transfer matrix (see e.g., \cite{mn/anal}, \cite{doikou}).
\be
t(\lambda) = t(-\lambda -\rho) \,,
\label{cross}
\ee provides further restrictions among the dressing functions i.e.,
\be
A_{3}(-\lambda - \rho) =  A_{1}(\lambda)\,, \non\\  A_{2}(\lambda) = A_{2}(-\lambda - \rho)\,.
\label{crosp}
\ee
The last two equations combined give
\be
A_{1}(\lambda)A_{1}(-\lambda) = 1\,.
\ee
 Moreover, for $\lambda= - i$ the $R$ matrix degenerates to a projector onto a three dimensional subspace. Thus, we can obtain another equation that involves $A_{1}(\lambda)$ and $A_{2}(\lambda)$ (see also \cite{reshe}), namely, 
\be
A_{2}(\lambda)A_{1}(\lambda+i)=A_{1}(\lambda+{i \over2})\,.
\ee
 Finally, we require $A_{2}(\lambda)$ to have the same poles with $A_{1}(\lambda)$ and $A_{3}(\lambda)$. Considering all the above conditions together we find that
\be
A_{1}(\lambda) =  \prod_{j=1}^{m} {\lambda+\lambda_{j}-{i\over2}\over \lambda+ \lambda_{j} +{i\over2}}{\lambda-\lambda_{j}-{i\over2}\over \lambda- \lambda_{j} +{i\over2}}\,, 
\ee
\be
A_{2}(\lambda) =  \prod_{j=1}^{m} {\lambda+\lambda_{j}+{3i\over2}\over \lambda+ \lambda_{j} +{i\over2}}{\lambda-\lambda_{j}+{3i\over2}\over \lambda- \lambda_{j} +{i\over2}}\non\\ \times {\lambda+\lambda_{j}\over \lambda+ \lambda_{j} +i}{\lambda-\lambda_{j}\over \lambda- \lambda_{j} +i}\,,
\ee
\be
A_{3}(\lambda) =  \prod_{j=1}^{m} {\lambda+\lambda_{j}+2i\over \lambda+ \lambda_{j} +i}{\lambda-\lambda_{j}+2i \over \lambda - \lambda_{j} +i}\,.
\ee
We can check that the above functions indeed satisfy all the necessary properties. Finally, the analyticity of the eigenvalues (the poles must vanish) provides the Bethe ansatz equations
\be
&&e_{1}(\lambda_{i})^{N}e_{-1}(2\lambda_{i})= \non\\ &&  -\prod_{j=1}^{m} e_{2}(\lambda_{i} - \lambda_{j})\ e_{2}(\lambda_{i} + \lambda_{j})
  \non\\ &\times& e_{-1}(\lambda_{i} - \lambda_{j})\ e_{-1}(\lambda_{i} + \lambda_{j})\,,
\label{BAE}
\ee
where we have defined $e_{n}(\lambda)$ as
\be
e_{n}(\lambda) = {\lambda + {in \over 2} \over \lambda - {in \over 2}}\,.
\label{DEF}
\ee
Notice that we obtain a completely new set of Bethe equations starting with the known $SU(3)$ invariant $R$ matrix. 
Furthermore, the result can be probably generalized for the spin chain constructed by the $SU(\n)$ invariant $R$ matrix. We expect a reduced symmetry for the general case as well.

\section{Discussion}
 
We constructed a quantum spin chain with ``soliton non-preserving'' boundary conditions. We used the symmetry of the model, the crossing symmetry and the fusion of the transfer matrix to find the spectrum of the transfer matrix, and we also deduced the Bethe ansatz equations (\ref{BAE}) via the analytical Bethe ansatz method. It would be of great interest to study the trigonometric case. Hopefully, one can find diagonal solutions for the $K$ matrices and solve the trigonometric open spin chain. The interesting aspect for the trigonometric case is that one can possibly relate the lattice model with some boundary field theory. Indeed, we know that e.g., the critical periodic $A_{\n-1}^{(1)}$ spin chain can be regarded as a discretisation of the corresponding affine Toda field theory \cite{zinn}. Finally, one can presumably generalize the above construction using any $SU(\n)$ invariant $R$ matrix. We hope to report on these issues in a future work \cite{new}.

\acknowledgments
This work was presented in the ``Non-perurbative quantum effects 2000'' TMR meeting in Paris. I am grateful to E. Corrigan, G.W. Delius, and R.I. Nepomechie for helpful discussions. This work was supported by the European Commission under the TMR Network ``Integrability, non-perturbative effects, and symmetry in quantum field theory'', number FMRX-CT96-0012.

\end{document}